\documentclass[11pt]{article}
\usepackage{amssymb, epsfig, amsmath, pifont, verbatim, graphicx,mathrsfs}
\graphicspath{{./}{../graphs/}{~/papers/reports/graphs/}}
\makeatletter
\renewcommand*\@fnsymbol[1]{\the#1}
\makeatother
\newcommand{\ud}{\mathrm{d}}
\title{Quantifying Chaos: A tale of two maps}
\author{R. L. Machete\footnote{Corresponding author: {\tt r.l.machete@reading.ac.uk}, tel: +44(0)118 378 6378}\\\\{\footnotesize Dept. of Mathematics and Statistics, P. O. Box 220, Reading, RG6 6AX, UK}}
\date{}
\newtheorem{proposition}{Proposition}
\begin{document}
\maketitle
\setcounter{footnote}{1}
\begin{abstract}
In many applications, there is a desire to determine if the dynamics of interest are chaotic or not. Since positive Lyapunov exponents are a signature for chaos, they are often used to determine this. Reliable estimates of Lyapunov exponents should demonstrate evidence of convergence; but literature abounds in which this evidence lacks. This paper presents two maps through which it highlights the importance of providing evidence of convergence of Lyapunov exponent estimates. The results suggest cautious conclusions when confronted with real data. Moreover, the maps are interesting in their own right.
\end{abstract}
\section{Introduction}
Computations of Lyapunov exponents continue to play a significant role in the study of nonlinear systems. The computations are often numerical because most nonlinear systems prove analytically intractable. In the case of real systems, there may be no mathematical model that adequately describes the system; hence one is often confronted with a limited amount of data from which to estimate Lyapunov exponents. Finite observations may afford computations of finite time Lyapunov exponents, but not global Lyapunov exponents. Unlike global Lyapunov exponents, finite time Lyapunov exponents depend on the initial conditions~\cite{er-85}. This limits their value in characterising the underlying system.

Fortunately, \cite{ose-abs} provides a theorem that guarantees convergence of finite time Lyapunov exponents to the global Lyapunov exponents. The hypothesis of the theorem (also found in~\cite{er-85}) requires the underlying system to have an invariant measure. It behoves us, therefore, to satisfy ourselves that the system of interest has an invariant measure and that our estimates converge to the global Lyapunov exponent. Yet examples abound in which Lyapunov exponent estimates are provided with neither of these two requirements confirmed (e.g. \cite{zie-smi,zha-09,mil-61,mar-10,bru-09,rei-08,ube-09}). Some have emphasised the need to provide confidence limits along with the estimates~\cite{gen-96,zie-jur}, but confidence limits are valuable only when there is evidence of convergence.

This paper presents two maps through which it highlights the importance of demonstrating evidence of convergence of Lyapunov exponent estimates. One map has a parameter that takes non-negative integer values. For all possible values of this parameter, we can compute the invariant measure in closed form. The other map presents computational challenges, despite its illusive simplicity. The two maps are presented in \S~\ref{sec:two}, with computations of Lyapunov exponents. We conclude with a discussion and summary of the results in \S~\ref{sec:disc}. Lyapunov Exponents are briefly discussed in the following section.
\section{Lyapunov Exponents}
Lyapunov exponents measure the average rate of separation of two trajectories that are initially infinitely close to each other. Consider a map
\begin{equation*}
x_{t+1}=F(x_{t}),
\end{equation*}
where $x_t\in R^m$. For an initial state $x_0$, the dynamics of its small perturbation, $\delta x_0$, are governed by the linear propagator, $\mathcal{M}(x_0,N)$, which is a product of Jacobians so that $\mathcal{M}(x_0,N)=DF(x_{N-1})\cdots DF(x_1)DF(x_0)$. The finite time average separation/growth rate of two initially nearby trajectories is then given by
\begin{equation*}
\lambda_N=\frac{1}{N}\log||\mathcal{M}\delta x_0||.
\end{equation*}
Oseledec~\cite{ose-abs} provides a theorem that guarantees that the limit $\lim_{N\rightarrow\infty}\lambda_N$ is unique. Denote this limit by $\Lambda$. If $\delta x_0$ is a member of the right singular vectors of $\mathcal{M}$, then $\lambda_N$ is a finite time Lyapunov exponent and $\Lambda$ is a global Lyapunov exponent.

It has been noted that $m>1$ corresponds to non-commuting matrices~\cite{er-85,zie-jur}. On the other hand, when $m=1$, we take logs of positive scalars so that the ergodic theorem may be used to yield
\begin{equation}
\Lambda=\int\rho(x)\log|F'(x)|\ud x,
\label{inv:lya}
\end{equation}
where $\rho(x)$ is the invariant distribution of the dynamics. If a dynamical system settles onto an invariant distribution, then the effect of transients on dynamical invariants averages out. Moreover, equation~(\ref{inv:lya}) provides an alternative to computing Lyapunov exponents when $m=1$; hence, simple bootstrap resampling techniques may be applied in a straight forward way to make Lyapunov exponent estimates, with no need for block resampling approaches suggested by Ziehmann et al.~\cite{zie-jur}. Given a set of observations, bootstrapping is accomplished by repeatedly sampling randomly from the data with replacement to estimate the statistic of interest~\cite{zie-jur}. This would give a distribution of the statistic of interest, which in this case is $\lambda_N$.

Since, for a given $N$, finite time Lyapunov exponents are functions of the initial states, it is useful to report values corresponding to a distribution of initial states to determine convergence. For an assessment of convergence of finite time Lyapunov exponents, one can sample the initial states from the invariant distribution, $\rho(x)$. In numerical computations, the invariant distribution is often not accessible in closed form. Nonetheless one can iterate forward an initial distribution $\rho_0(x)$ so that after $N$ iterations $\rho_N(x)$ is an estimate of the invariant distribution. One would then sample initial states from $\rho_N(x)$ to estimate $\lambda_N$'s. The aim here is to sample the initial states according to the invariant measure, if it exists. As a consequence of Oseledec's theorem~\cite{ose-abs}, the distribution of the $\lambda_N$'s will converge to a delta distribution centred at $\Lambda$ as $N\rightarrow\infty$. 

When one is faced with real data, the approach of the previous paragraph cannot be used unless there is a reliable mathematical model of the system. Nonetheless, one may use bootstrap approaches suggested in~\cite{zie-jur} for various values of $N$ to assess convergence in the distribution of the $\lambda_N$'s. 
\section{The Two Maps}
\label{sec:two}
In this section, two novel maps with contrasting behaviours are considered. Analytic computations are complemented with numerical results and the importance of establishing convergence is highlighted. 
\subsection{Infinitely Piece-wise Linear Map}
\begin{figure}[!t]
\centering
\includegraphics[height=7.5cm,width=7.5cm]{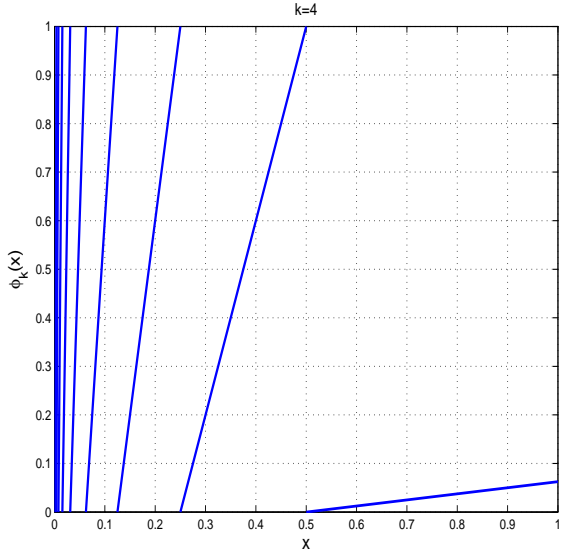}
\caption[Map Graph]{\parbox[t]{0.9\linewidth}{\small (Left) Graph of the Infinitely piece-wise linear map given in equation~(\ref{eqn:inf}).}}
\label{fig:piece}
\end{figure}

Let us consider the infinitely piece-wise linear map
\begin{equation}
\phi_k(x)=\left\{\begin{array}{ll} \frac{2}{2^k}(x-\frac{1}{2}),& x\in[\frac{1}{2},1],\\\\2^i(x-\frac{1}{2^i}), & x\in[2^{-i},2^{1-i}),\quad i=2,3,\ldots\end{array}\right.
\label{eqn:inf}
\end{equation}
defined for fixed $k\in\{0,1,\ldots,\infty\}$. A graph of the map corresponding to $k=4$ is shown in figure~\ref{fig:piece}. 
The Lyapunov exponents for this map may be sought either analytically or numerically. For a given value of $k$, an exact knowledge of the invariant measure of this map is sufficient for one to determine the corresponding global Lyapunov exponent. 
\subsubsection{Invariant Densities}
Denote the invariant density corresponding to a given value of $k$ by $\rho^{(k)}(x)$ and the associated invariant measure~\footnote{See~\cite{er-85} for details of invariant measures.} by $\mu^{(k)}$. For $k=0$, the invariant distribution of this map is the uniform distribution $U[0,1]$. That is
\begin{equation*}
\rho^{(0)}(x)=\left\{\begin{array}{ll} 1,& x\in[0,1],\\0, &\mbox{otherwise.}\end{array}\right.
\end{equation*}
To see this, we note that the invariant density must satisfy the Perron-Frobenius equation~\cite{las-mac,las-mac2}, whence
\begin{equation*}
\rho^{(0)}(x)=\sum_{i=1}^{\infty}\frac{\rho^{(0)}[2^{-i}(x+1)]}{2^i}.
\end{equation*}
It is then evident that the uniform distribution satisfies the above equation identically. When $k>0$, the following propositions hold and their proofs are given in appendix~\ref{sec:app1}:
\begin{proposition} For $k=1$, the invariant density of the piecewise linear map is
\begin{equation*}
\rho^{(1)}(x)=\left\{
\begin{array}{ll}
\frac{4}{3}, & x\in \left[0,\frac{1}{2}\right),\\\\
\frac{2}{3}, & x\in\left[\frac{1}{2},1\right],
\end{array}
\right.
\end{equation*}
\end{proposition}
\begin{proposition}
For any $k\ge2$, the invariant density of the infinitely piecewise linear map is
\begin{equation*}
\rho^{(k)}(x)=\left\{
\begin{array}{ll}
\frac{2^k+2}{3}, & x\in\left[0,\frac{1}{2^k}\right),\\\\
\frac{2}{3}, &x\in\left[\frac{1}{2^k},\frac{1}{2}\right),\\\\
\frac{2}{3}, &x\in\left[\frac{1}{2},1\right].
\end{array}
\right.
\end{equation*}
\end{proposition}
\subsubsection{Global Lyapunov Exponents}
The Lyapunov exponents of the map are given by
\begin{equation*}
\Lambda^{(k)}=\int_0^1\rho^{(k)}(x)\log_2\phi_k^{'}(x)\ud x,
\end{equation*}
where the derivative is taken only at points of continuity, and is thus a weak derivative. We then state the following proposition whose proof is given in appendix~\ref{sec:app2}:
\begin{proposition} For any $k\in\{0,1,2,\ldots\}$, the global Lyapunov exponent of the infinitely piecewise linear map is
$$\Lambda^{(k)}=2.$$
\end{proposition}
\subsubsection{Numerical Results}
\begin{figure}[!t]
%\hspace{-2.7cm}
\hbox{
\includegraphics[height=7.5cm,width=7.5cm]{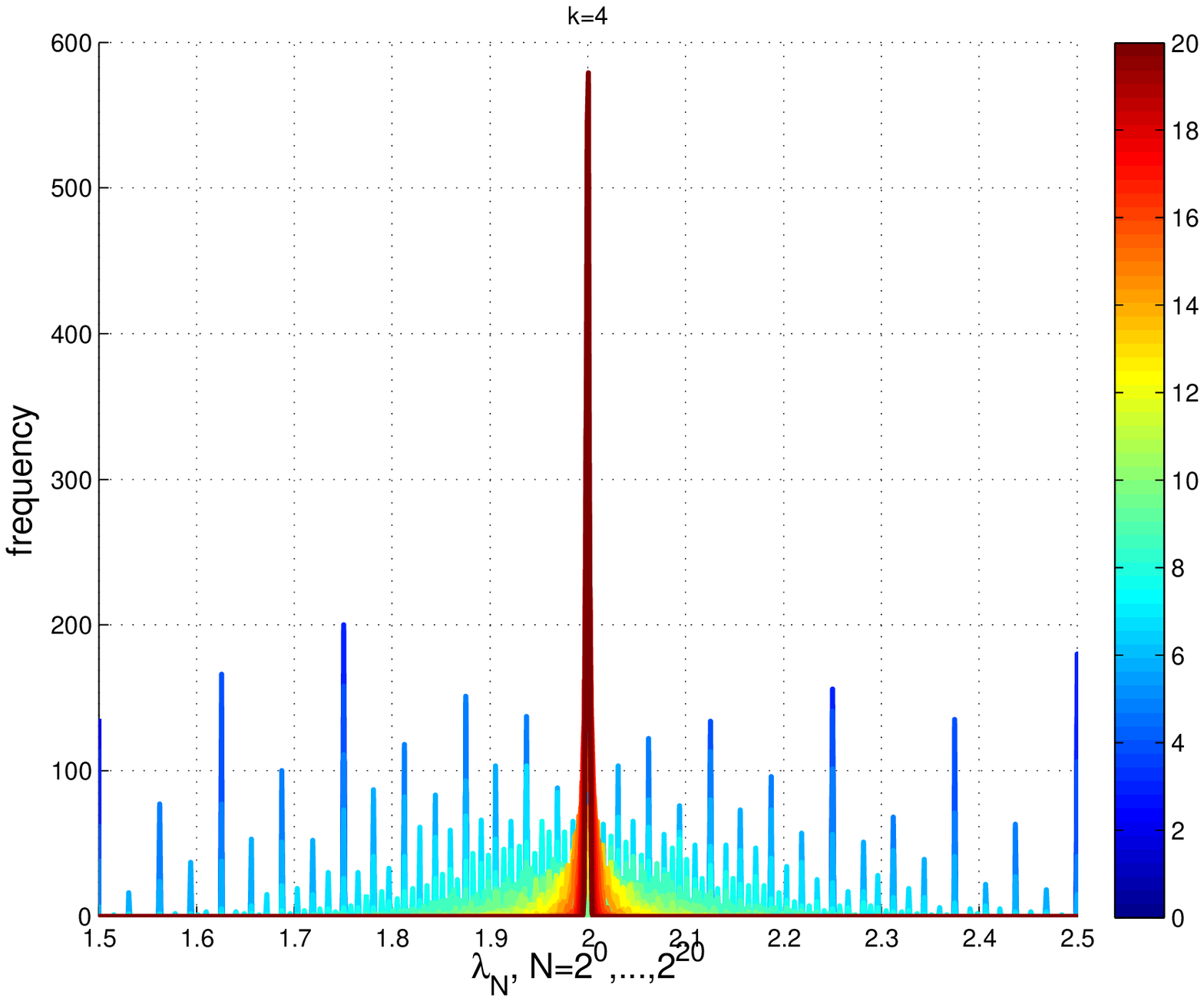}
\includegraphics[height=7.5cm,width=7.5cm]{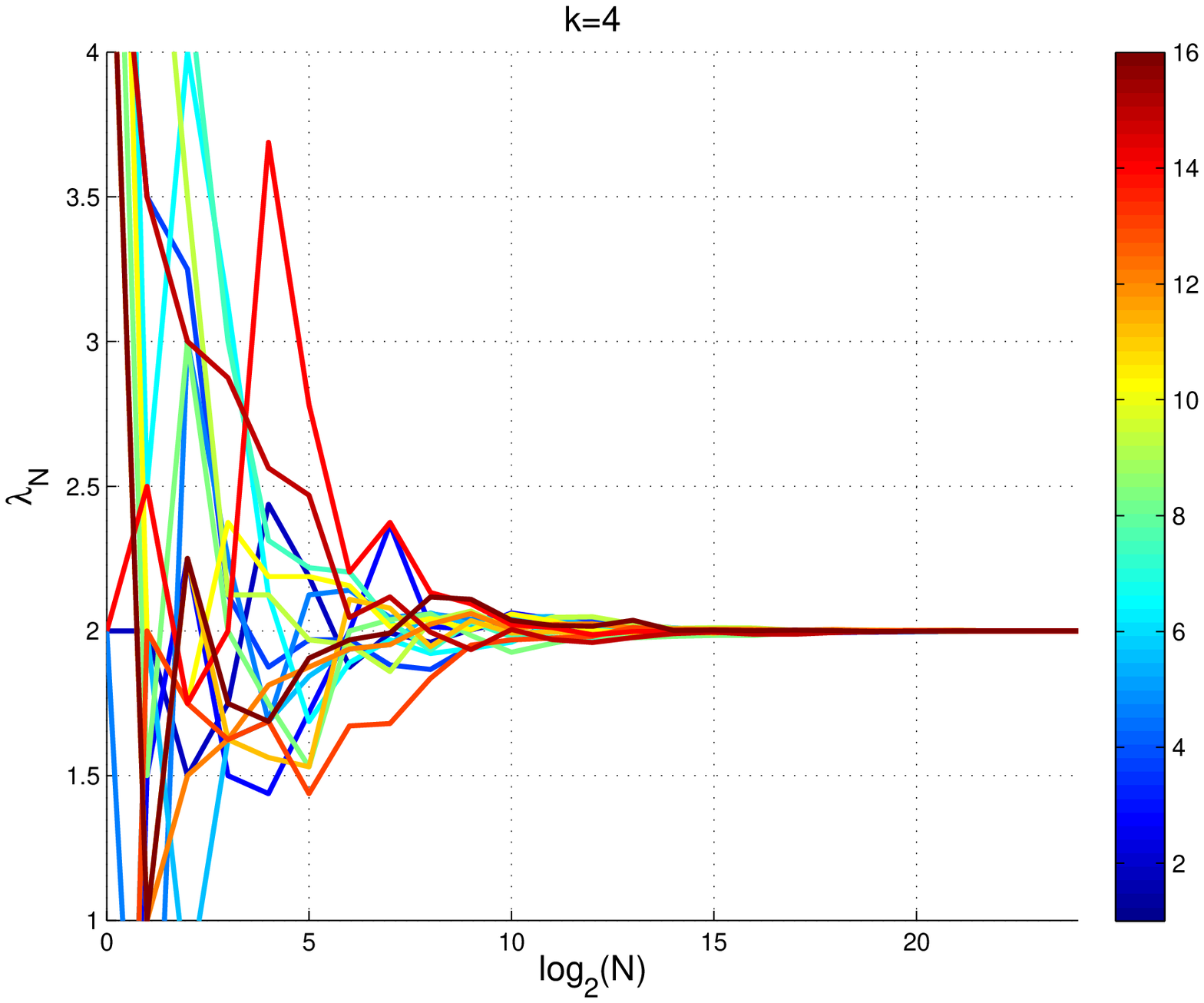}
}
\caption[Map Graph]{\parbox[t]{0.9\linewidth}{\small (Left) Distributions of finite time Lyapunov exponents for the infinitely piecewise map in~(\ref{eqn:inf}). The different distributions correspond to different time lengths of trajectories according to the colour bar. (Right) Finite time Lyapunov exponents from $2^4$ as a function of time, $N=2^n$. The different lines correspond to different initial conditions.}}
\label{inf:lyap}
\end{figure}

Numerical estimates of global Lyapunov exponents may be obtained in either of two ways. One approach is to compute finite time Lyapunov exponents along trajectories from different initial conditions. The distribution of finite time Lyapunov exponents should converge to a delta distribution as the length of the trajectories increases. For the infinitely piecewise linear map in equation~(\ref{eqn:inf}) with $k=4$, we considered trajectories of length $N=2^n$ starting from $2^{10}$ initial conditions. We varied $n$ from 0 to 20 and the resulting distributions are shown in figure~\ref{inf:lyap}. Notice that with every doubling of the length of trajectories, the distribution of Lyapunov exponents converges to a delta distribution that is centred at 2.

Alternatively, we can plot finite time Lyapunov exponents corresponding to each initial condition as a function of trajectory length, $N$. These are also plotted in figure~\ref{inf:lyap} on the right hand side. All the lines corresponding to $2^4$ initial conditions clearly converge to a value around 2.

\begin{figure}[t]
%\hspace{-2.7cm}
\hbox{
\includegraphics[height=7.5cm,width=7.5cm]{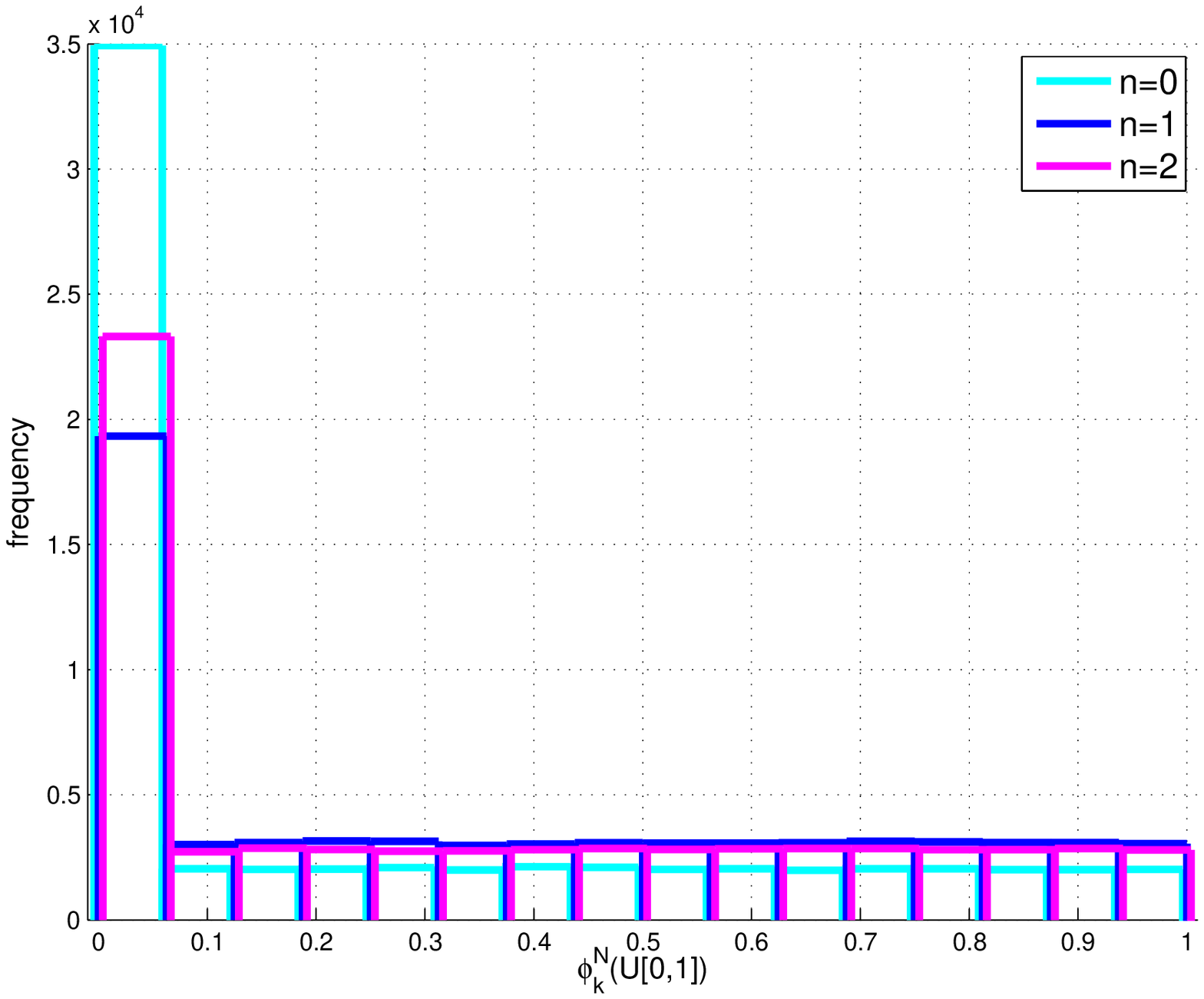}
\includegraphics[height=7.5cm,width=7.5cm]{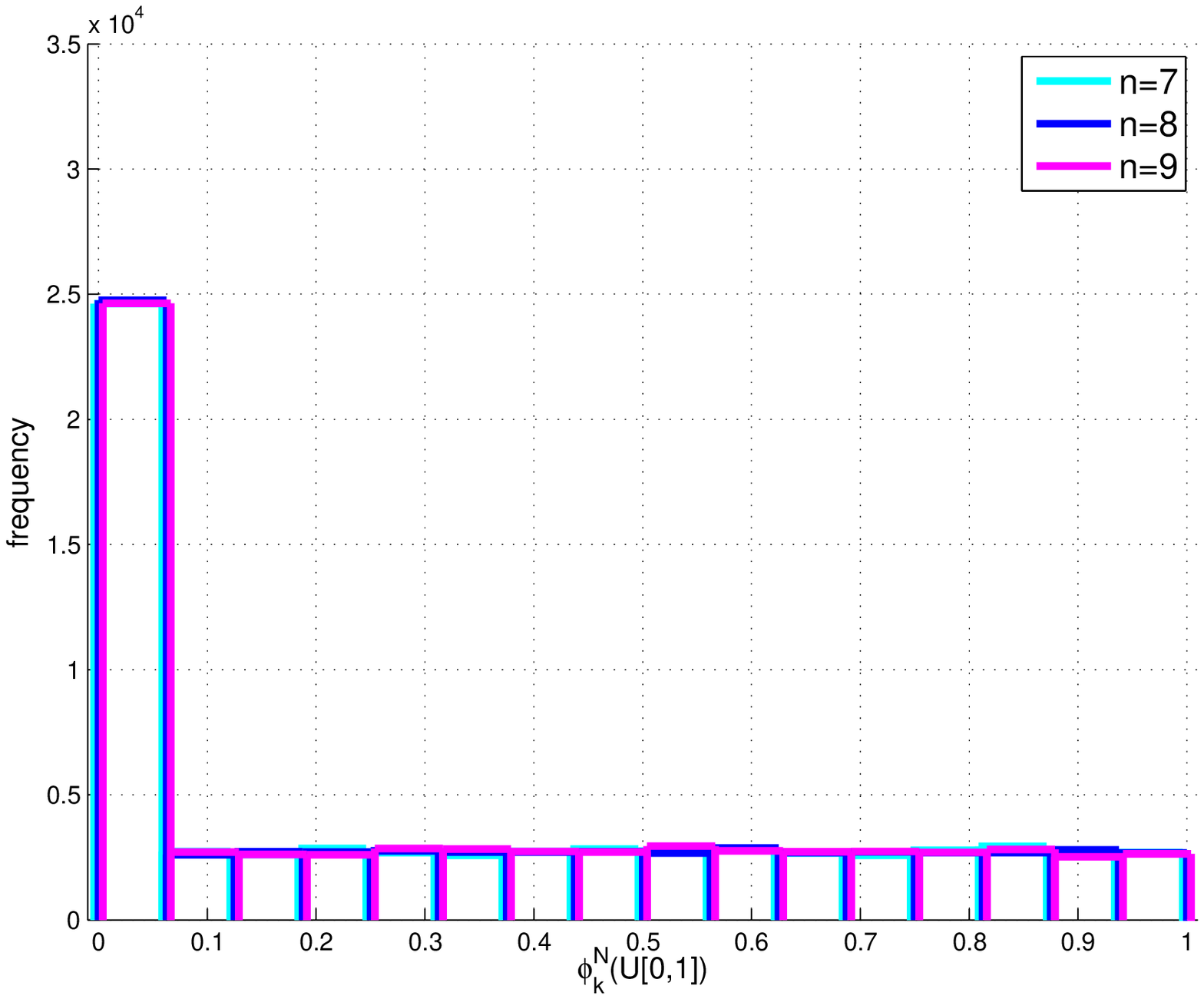}
}
\caption[Map Graph]{\parbox[t]{0.9\linewidth}{\small (Left) Evolution of $2^{16}$ points sampled from $U[0,1]$ under the infinitely piecewise map in~(\ref{eqn:inf}) with the number of iterations given by $N=2^n$. The histograms were obtained by using $2^{4}$ equally spaced bins on the unit interval.}}
\label{inf:dist}
\end{figure}

\begin{figure}[t]
%\hspace{-2.7cm}
\hbox{
\includegraphics[height=7.5cm,width=7.5cm]{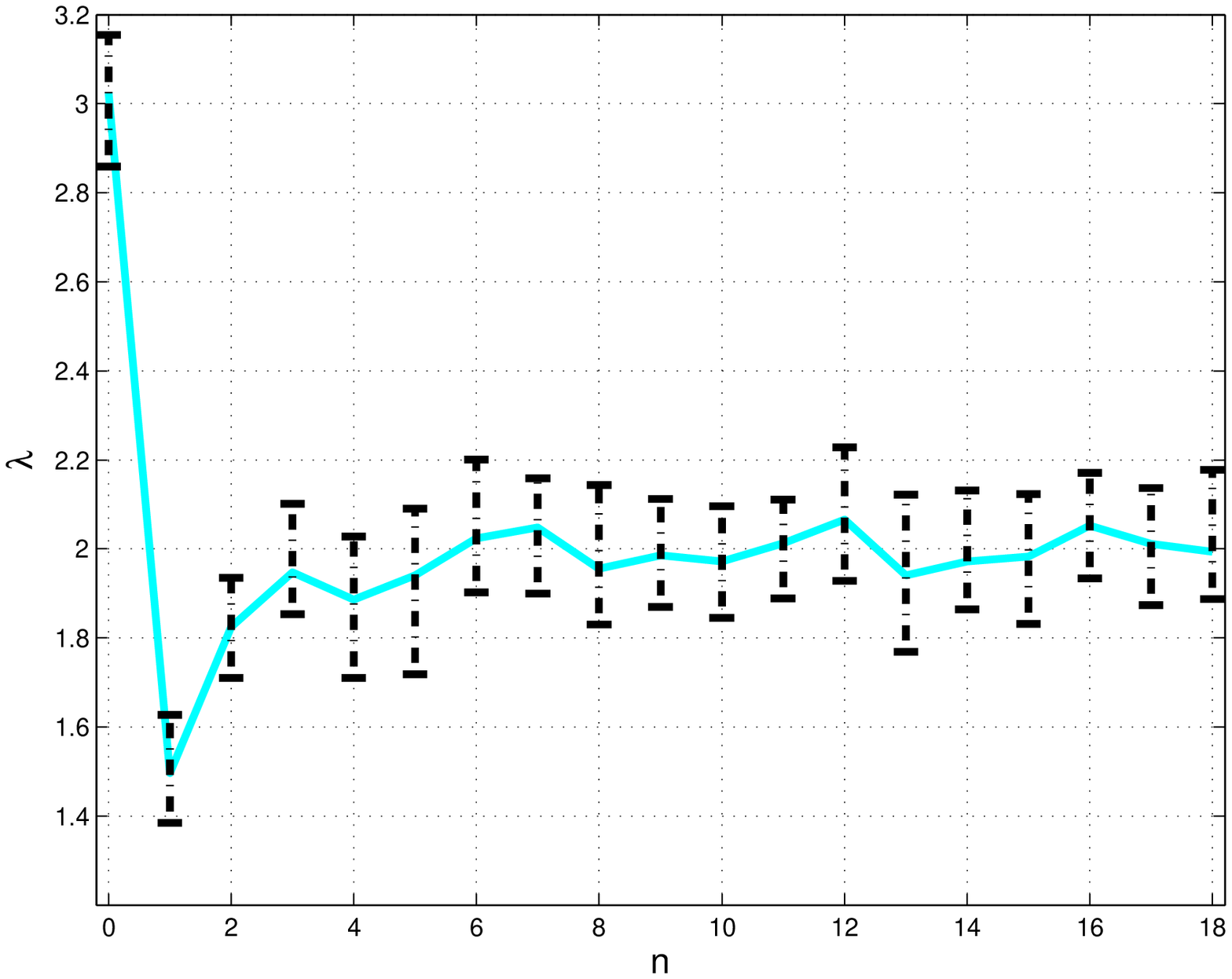}
\includegraphics[height=7.5cm,width=7.5cm]{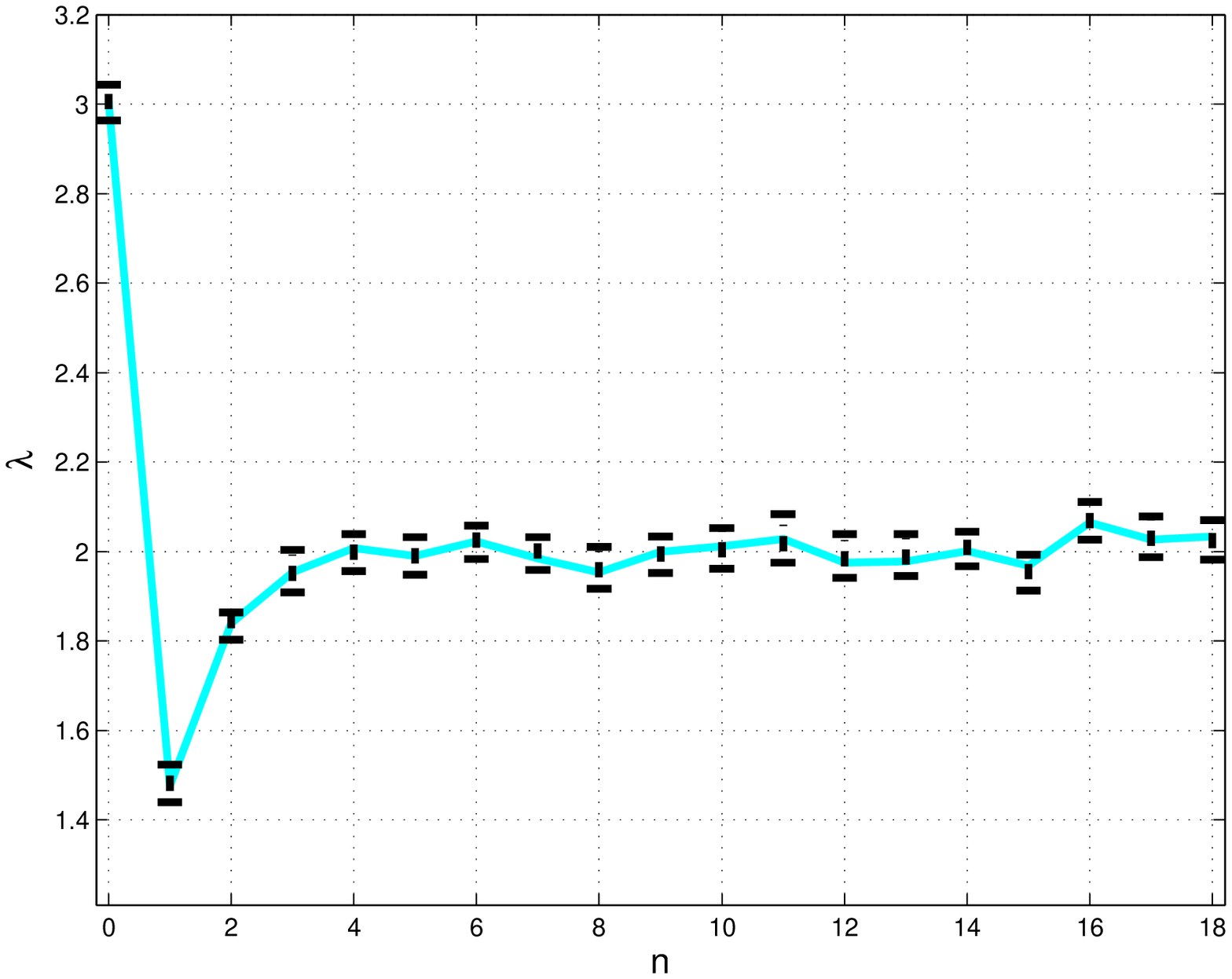}
}
\caption[Limiting Distribution]{\parbox[t]{0.9\linewidth}{\small Lyapunov exponent estimates from a distribution obtained by evolving forward $M=2^{13}$ (left) and $M=2^{16}$ (right) points sampled from $U[0,1]$ under the infinitely piece-wise linear map when $k=4$. Each distribution is thus $\phi_k^N(U[0,1])$, where $N=2^n$. In each graph, the solid line corresponds to the mean estimate and the dash-dotted lines correspond to bootstrap uncertainty estimates.}}
\label{dens:exp}
\end{figure}

Finally, we could compute the Lyapunov exponents by first estimating the invariant density. The invariant density may be estimated by evolving forward the standard uniform distribution $U[0,1]$ under the map $\phi_k(x)$. Again we considered $k=4$ and evolved forward $M$ points sampled from $U[0,1]$. A time series of distributions corresponding to $M=2^{16}$ is shown in figure~\ref{inf:dist}. The distribution appears to have converged to the invariant distribution after $N=2^n$, when $n=7$. Lyapunov exponent estimates corresponding to $M=2^{13}$ (left) and $M=2^{16}$ (right) are shown in figure~\ref{dens:exp}, where the solid line corresponds to the mean and the dash-dotted lines are bootstrap uncertainty estimates. Notice that the value of $\Lambda^{(k)}=2$ is generally within the confidence intervals for $n\ge4$. The confidence intervals on the left are relatively big. This is due to $M=2^{13}$ being too small to provide a good estimate of the invariant distribution. As may be noted from the right hand graphs on figure~\ref{inf:lyap}, convergence was obtained with more than $N=2^{15}$ points. It is, therefore, not surprising that when $M=2^{16}$ the confidence intervals are much smaller as is the case on the right hand graph of the same figure.


\subsection{Paradoxical Map}
\begin{figure}
\centering
\hbox{
\includegraphics[height=7.5cm,width=7.5cm]{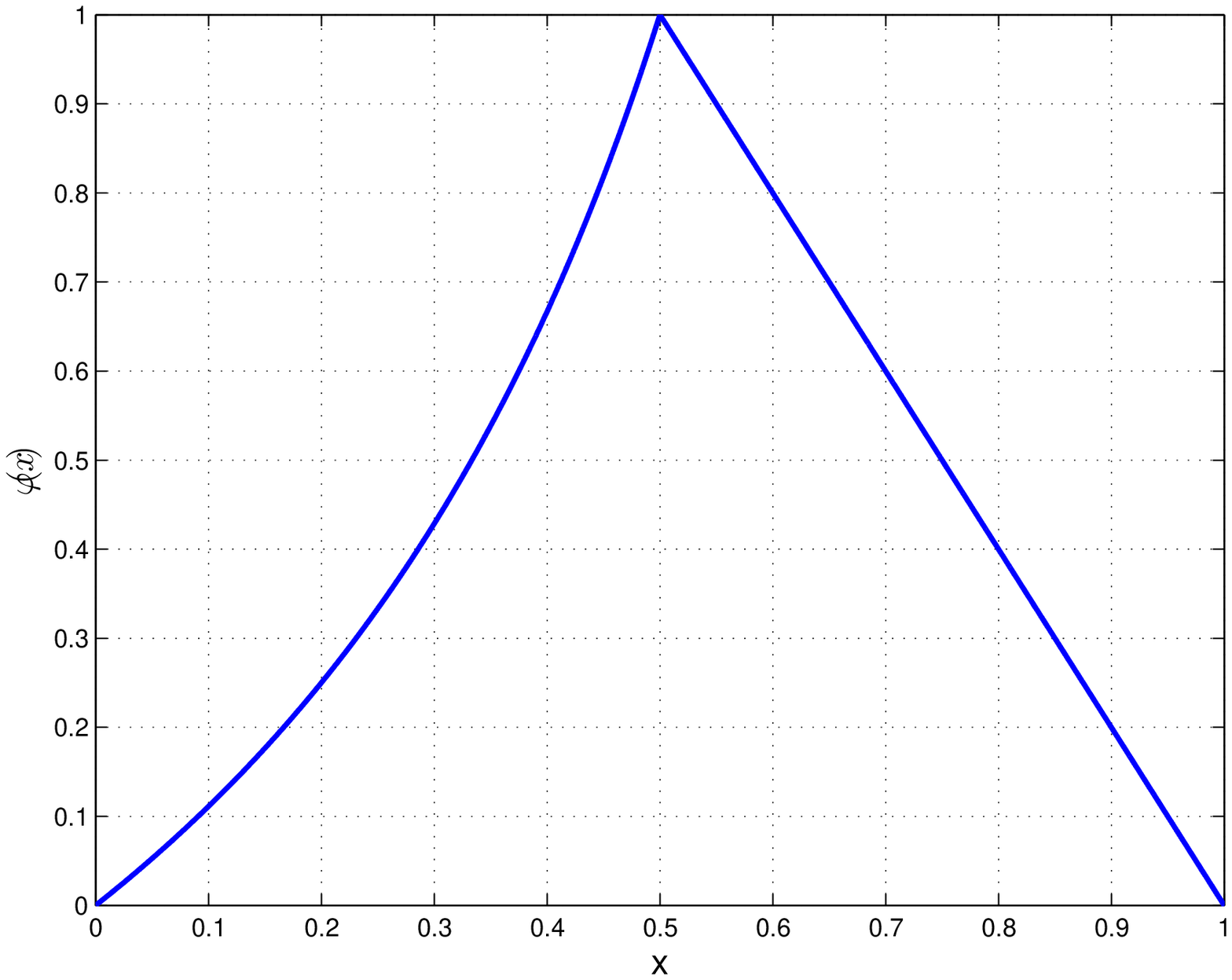}
\includegraphics[height=7.5cm,width=7.5cm]{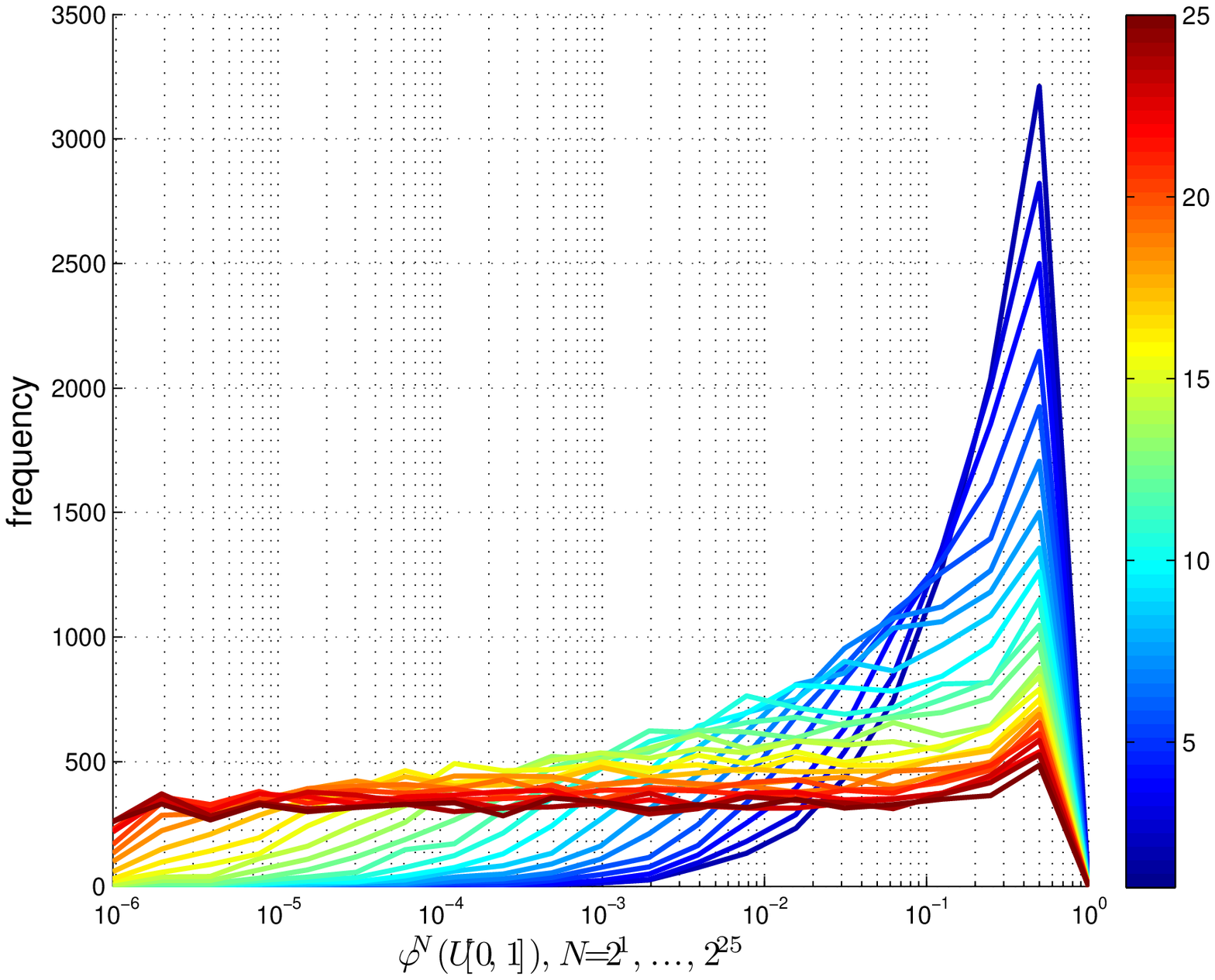}
}
\caption[Limiting Distribution]{\parbox[t]{0.9\linewidth}{\small (Left) Graph of the map $\varphi(x)$ given in equation~(\ref{eqn:par}). (Right) Evolution of $2^{15}$ initially uniformly distributed ensemble in $[0,1]$ under the map $\varphi(x)$. The bins for these histograms were logarithmically spaced to the base 2.}}
\label{fig:liapr1}
\end{figure}

The map considered here is
\begin{equation}
\varphi(x)=\left\{\begin{array}{ll}
\frac{x}{1-x}, & x\in(0,1/2),\\\\
2(1-x), & x\in(1/2,1).
\end{array}\right.
\label{eqn:par}
\end{equation}
Its graph is given in figure~\ref{fig:liapr1} on the left panel. It has an unstable fixed point at the origin. The stability of the origin is determined using
\begin{equation*}
\varphi^{'}(x)=\frac{1}{(1-x)^2}.
\end{equation*}
Clearly, $\varphi^{'}(x)>1$ for any $x>0$. However, when $x$ is close to zero then $\varphi^{'}(x)$ is close to~1. Hence the origin is a weak repeller as pointed out in~\cite{las-mac}. Indeed it is this property that makes the dynamics of this map paradoxical as was found in~\cite{las-mac,las-mac2} on a similar map that differs only by the linear part. If $\varphi^{j}(x)\in(0,1/2)$ for all $j=0,1,\ldots,n-1$, then
\begin{equation*}
\varphi^n(x)=\frac{x}{1-nx},
\end{equation*}
where $\varphi^0(x)=x$. On the hand, if $\varphi^{j}(x)\in(1/2,1)$ for all $j=0,1,\ldots,n-1$, then
\begin{equation*}
\varphi^n(x)=2\left[\frac{1-(-2)^n}{3}-(-2)^{n-1}x\right].
\end{equation*}
It is, therefore, evident that with probability one trajectories with initial conditions sampled according to the Lebesgue measure on $[0,1]$ do not converge to the origin. However, it can be shown that iterates of distributions that are initially uniform converge to a delta distribution centred at the origin. These two contrasting behaviours of trajectories and densities are paradoxical and are due to the weak repeller at the origin.

Numerically, the invariant density of this map may be determined by evolving forward in time the uniform distribution $U[0,1]$. Iterating forward an initial ensemble of $2^{15}$ points sampled from $U[0,1]$ yields the graphs shown in figure~\ref{fig:liapr1} on the right panel. From the graphs, $\varphi^N(U[0,1])$ appears to have nearly converged to the invariant density when $N=2^{24}$. If that is indeed the case, it follows that one can compute estimates of the global Lyapunov exponents using points sampled from $\varphi^N(U[0,1])$ when $N=2^{25}$ via equation~(\ref{inv:lya}). Unfortunately, the dynamics are trapped into very long periodic orbits after about $N=2^{25}$ iterations, the most dominant period being of the order of $2^{29}$. %A summary of the results of iterating forward $U[0,1]$ is given in figure~\ref{fig:orbits}.
%\begin{figure}[!t]
%\hspace{-2.0cm}
%\vbox{
%\hbox{
%\includegraphics[height=7.5cm,width=7.5cm]{../graphs/server_orbit_frequency.eps}
%\includegraphics[height=7.5cm,width=7.5cm]{../graphs/server_orbit_reach_time.eps}
%}
%\hbox{
%\includegraphics[height=7.5cm,width=7.5cm]{../graphs/server_orbit_exponent.eps}
%\includegraphics[height=7.5cm,width=7.5cm]{../graphs/server_orbit_period.eps}
%}
%}
%\caption[Long term dynamics]{\parbox[t]{0.9\linewidth}{\small Details of the Periodic orbits that the maps settle on for any randomly selected initial condition in $U[0,1]$. Each periodic orbit is identified by its corresponding minimum.}}
%\label{fig:orbits}
%\end{figure}

\begin{figure}[t]
\centering
\hbox{
\includegraphics[height=7.5cm,width=7.5cm]{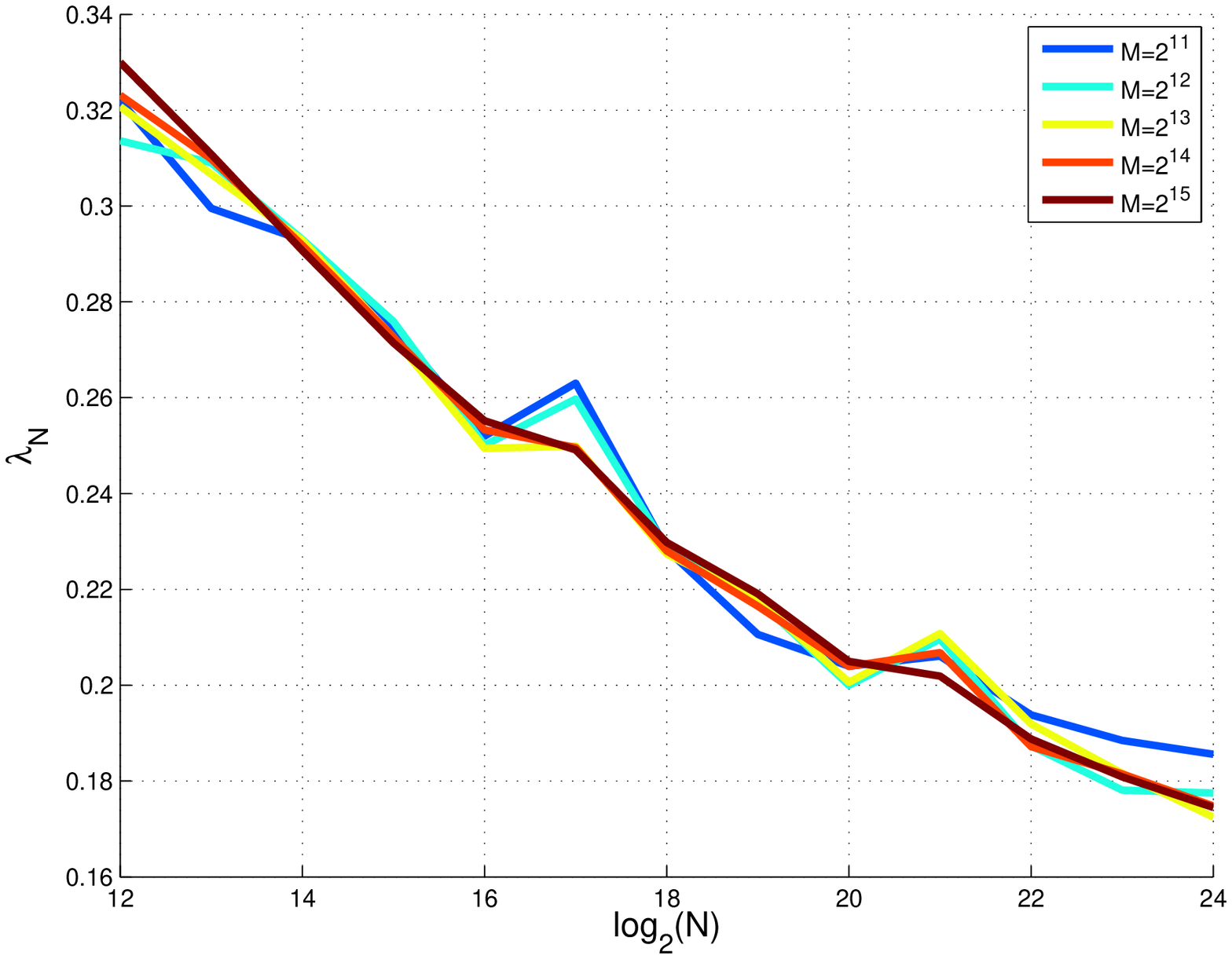}
\includegraphics[height=7.5cm,width=7.5cm]{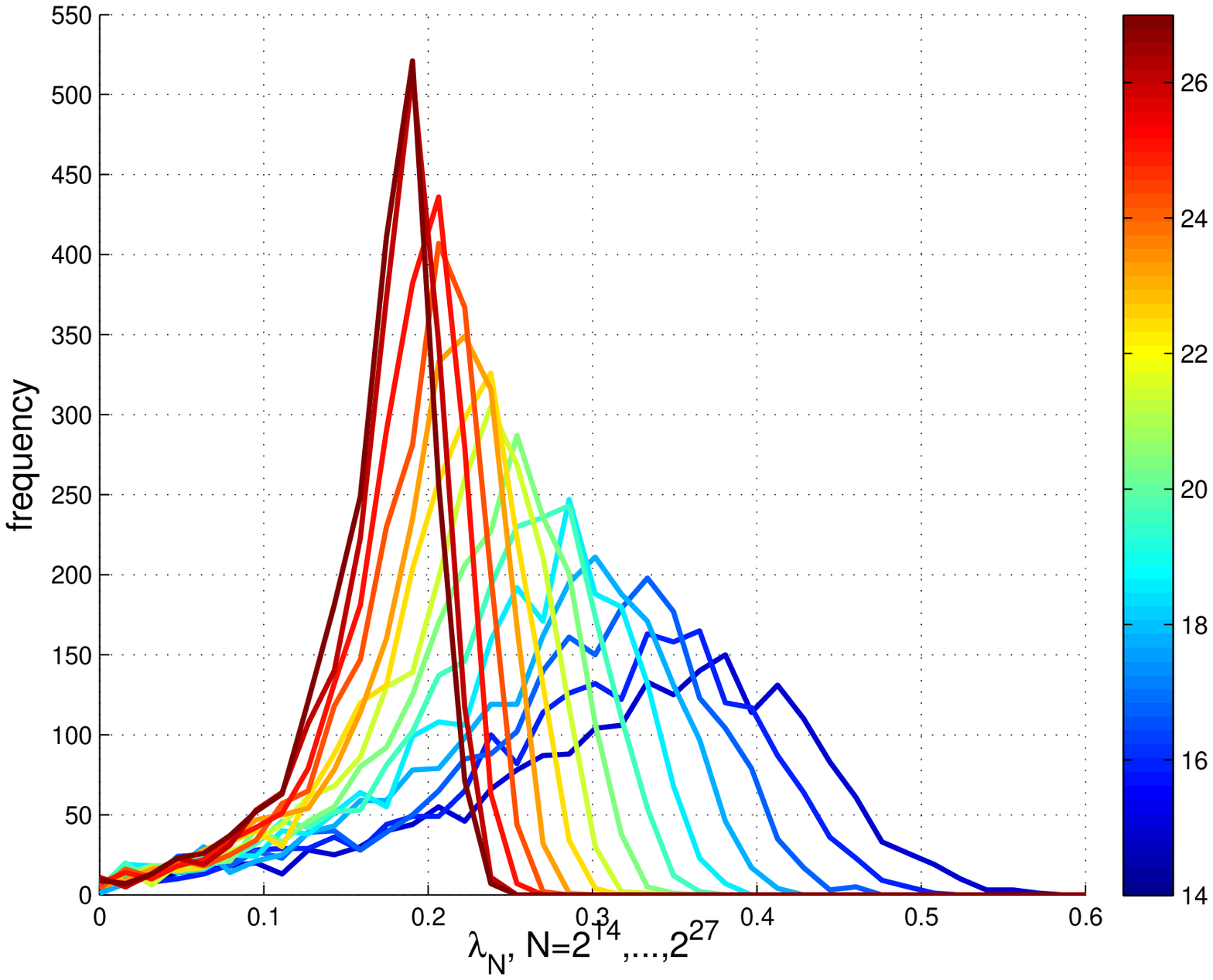}
}
\caption[Limiting Distribution]{\parbox[t]{0.8\linewidth}{\small (left)Graphs of finite time Lyapunov exponents versus trajectory length. (right) Distributions of Lyapunov Exponents. Each distribution is obtained by computing finite time Lyapunov exponents along trajectories whose initial conditions are sampled from $U[0,1]$. }}
\label{fig:few}
\end{figure}

Distributions of finite time Lyapunov exponents for varying $N$ are shown in figure~\ref{fig:few} on the right panel. The initial conditions of trajectories used to estimate each $\lambda_N$ were sampled from $\varphi^N(U[0,1])$. Whereas it seems the distribution of Lyapunov exponents has converged when $N=2^{25}$, the distribution converged to is not a delta distribution. In fact, if iterations are continued further, the distribution ultimately becomes tri-modal, each mode corresponding to each periodic orbit that the dynamics settle onto. If one had only a finite sample of data without a knowledge of the data-generating process, as is typical in practical situations, they could mistakenly attribute the bell shape to uncertainty.

Finally, graphs of $\lambda_N$, computed along trajectories starting from different initial conditions, versus $\log N$ indicated that there is indeed no convergence~(see figure~\ref{fig:few} on the left panel). Each line on the graph corresponds to a number of initial conditions, $M$. This raises suspicion over Lyapunov exponent estimates obtained from a single initial condition. An example of where this was done is~\cite{zha-09}, albeit using real data. 
\section{Discussion}
\label{sec:disc}
The aim of this paper was to highlight the importance of demonstrating convergence of finite time Lyapunov exponents estimates of the global Lyapunov exponent. Two contrasting maps were thus presented, one of which its dynamical properties are known accurately, and the other not. The former map has a parameter that takes non-negative integer values. As the sole parameter is varied across all non-negative integer values, the global Lyapunov exponent remains fixed at $\Lambda=2$. At a sample of parameter values, numerical computations confirmed a fast convergence of distributions of finite time Lyapunov exponents to a delta distribution. The centre of the delta distributions coincided with the analytically found value of 2. 

The other map was more challenging. Detailed insights into its dynamics were sought numerically. Numerical distributions of finite time Lyapunov exponent estimates did not converge to a delta distribution; even with the use of trajectories as long as $2^{25}$. If the length of trajectories is increased, the dynamics ultimately fall onto very long periodic orbits due to finite machine precision. The lack of convergence of distributions of finite time Lyapunov exponents to a delta distribution should alarm us against reporting the mode as an estimate of the global Lyapunov exponent. Indeed any arising confidence limits would equally be nonsensical. In fact, following analytic considerations of a similar map in~\cite{las-mac,las-mac2} would indicate that the map has no invariant distribution. Hence, for a randomly selected initial condition, the dynamics are transient with probability one.  

The failure of distributions of finite time Lyapunov exponents to converge to a delta distribution is a caution against reporting estimates of Lyapunov exponents and confidence limits. In numerical computations, convergence of the distributions to a delta distribution is achievable if the dynamics settle onto an invariant distribution before being trapped onto a numerical periodic orbit. A higher machine precision may be necessary to ensure this. When using data from a real nonlinear system, stationarity has to be established first. Convergence of distributions of finite time Lyapunov estimates should then be verified by progressively increasing the length of trajectories from which the estimates are made. Block resampling approaches suggested in~\cite{zie-jur} may help provide the distributions that are used to assess convergence to a delta distribution.
\section*{Acknowledgements}
I would like to thank Prof L. A. Smith for useful discussions and contributions. I am grateful for comments from two anonymous reviewers that helped improve the manuscript. This work was supported by the RCUK Digital Economy Programme via EPSRC grant EP/G065802/1 The Horizon Digital Economy Hub.
\bibliographystyle{elsarticle-num}

%\bibliography{refs}
\appendix
\numberwithin{equation}{section}
\section{Proofs of Propositions 1 and 2}
\label{sec:app1}
{\bf Proof of Proposition 1}: When $k=1$, we note that the invariant density is piecewise constant with
\begin{equation*}
\rho^{(1)}(x)=\left\{
\begin{array}{ll}
c_1, & x\in A_1=\left[0,\frac{1}{2}\right),\\\\
c_2, & x\in A_2=\left[\frac{1}{2},1\right],
\end{array}
\right.
\end{equation*}
where $c_1$ and $c_2$ are constants.  In order to determine the constants $c_1$ and $c_2$, we note that by definition an invariant measure satisfies the property $\mu^{(k)}(A)=\mu^{(k)}\{\phi_k^{-n}(A)\}$~\cite{er-85}, where $\phi_k^{-n}(A)$ is the set of all the points that are mapped onto $A$ after $n$ iterations of the map $\phi_k(x)$. It thus follows that
\begin{equation}
\mu^{(1)}(A_1)=\frac{1}{2}\mu^{(1)}(A_1)+\mu^{(1)}(A_2).
\label{sec1:eq6}
\end{equation}
For a probability measure, it is also true that
\begin{equation}
\mu^{(1)}(A_1)+\mu^{(1)}(A_2)=1.
\label{sec1:eq7}
\end{equation}
Solving~(\ref{sec1:eq6}) and~(\ref{sec1:eq7}) yields
\begin{equation*}
\mu^{(1)}(A_1)=\frac{2}{3},\quad\mu^{(1)}(A_2)=\frac{1}{3}.
\end{equation*}
Hence $c_1=4/3$ and $c_2=2/3$.\\\\
{\bf Proof of Proposition 2}: Let us now consider the general case of $k\ge2$. In this case, the invariant density is constant on three intervals. That is
\begin{equation*}
\rho^{(k)}(x)=\left\{
\begin{array}{ll}
c_1^{(k)}, & x\in A_1^{(k)}=\left[0,\frac{1}{2^k}\right),\\\\
c_2^{(k)}, &x\in A_2^{(k)}=\left[\frac{1}{2^k},\frac{1}{2}\right),\\\\
c_3^{(k)}, &x\in A_3^{(k)}=\left[\frac{1}{2},1\right].
\end{array}
\right.
%\label{sec1:eq8}
\end{equation*}
Again, we use the underlying invariant measure to compute the $c_j^{(k)}$'s. If we let $\mu_j^{(k)}=\mu^{(k)}(A_j)$, then
\begin{eqnarray}
\label{sec1:eq9}
\mu_1^{(k)}&=&\frac{1}{2^k}\left[\mu_1^{(k)}+\mu_2^{(k)}\right]+\mu_3^{(k)},\\\nonumber\\\label{sec1:eq10}
\mu_2^{(k)}&=&\left(\frac{1}{2}-\frac{1}{2^k}\right)\left[\mu_1^{(k)}+\mu_2^{(k)}\right],\\\nonumber\\
\mu_3^{(k)}&=&\frac{1}{2}\left[\mu_1^{(k)}+\mu_2^{(k)}\right],
\label{sec1:eq11}
\end{eqnarray}
 Notice that the sum of the first two equations yields the last equation. Therefore, we need another equation in order to determine the $\mu_j^{(k)}$'s, which is
\begin{equation}
\mu_1^{(k)}+\mu_2^{(k)}+\mu_3^{(k)}=1.
\label{sec1:eq12}
\end{equation}
Using any two of equations~(\ref{sec1:eq9}),~(\ref{sec1:eq10})~and~(\ref{sec1:eq11})  and equation~(\ref{sec1:eq12}) yields
\begin{equation*}
\mu_1^{(k)}=\frac{2^{k-1}+1}{3\cdot 2^{k-1}},\quad
\mu_2^{(k)}=\frac{2^{k-1}-1}{3\cdot 2^{k-1}},\quad
\mu_3^{(k)}=\frac{1}{3}.
%\label{sec1:eq13}
\end{equation*}
From these it follows that
\begin{equation*}
c_1^{(k)}=\frac{2^k+2}{3},\quad
c_2^{(k)}=\frac{2}{3},\quad
c_3^{(k)}=\frac{2}{3}.
%\label{sec1:eq14}
\end{equation*}
\section{Proof of Proposition 3}
\label{sec:app2}
The proof of this proposition considers each of the cases $k=0$, $k=1$ and $k\ge2$ separately. When $k=0$, we get
\begin{eqnarray*}
\Lambda^{(0)}&=&\int_0^1\rho^{(0)}\log_2\phi_0^{'}(x)\ud x\\\\
            &=&\int_0^1\log_2\phi_0^{'}(x)\ud x\\\\
            &=&\sum_{i=1}^{\infty}\int_{\frac{1}{2^i}}^{\frac{1}{2^{i-1}}}\log_2\phi_0^{'}(x)\ud x\\
           &=&\sum_{i=1}^{\infty}\frac{i}{2^i}\\\\
           &=&2.
\end{eqnarray*}
Similarly, for $k=1$ we get
\begin{eqnarray*}
\Lambda^{(1)}&=&\int_0^1\rho^{(1)}(x)\log_2\phi_1^{'}(x)\ud x\\\\
             &=&\frac{4}{3}\int_0^{1/2}\log_2\phi_1^{'}(x)\ud x+\frac{2}{3}\int_{1/2}^1\log_2\phi_1^{'}(x)\ud x\\
            &=&\frac{4}{3}\sum_{i=2}^{\infty}\frac{i}{2^i}+0\\
            &=&2.
\end{eqnarray*}
For general $k\ge2$,
\begin{eqnarray*}
\Lambda^{(k)}&=&\int_0^1\rho^{(k)}(x)\log_2\phi_k^{'}(x)\ud x\\\\
&=&c_1^{(k)}\int_0^{\frac{1}{2^k}}\log_2\phi_k^{'}(x)\ud x+c_2^{(k)}\int_{\frac{1}{2^k}}^{\frac{1}{2}}\log_2\phi_k^{'}(x)\ud x+c_3^{(k)}\int_{\frac{1}{2}}^{1}\log_2\phi_k^{'}(x)\ud x\\\\
&=&\frac{(2^k+2)}{3}\sum_{i=k+1}^{\infty}\frac{i}{2^i}+\frac{2}{3}\sum_{i=2}^k\frac{i}{2^i}+(1-k)\cdot\frac{2}{3}\cdot\frac{1}{2}\\
&=&\frac{(2^k+2)}{3}\frac{(k+2)}{2^k}+\frac{2}{3}\cdot\frac{1}{2^k}\left[3\cdot 2^{k-1}-(k+2)\right]+\frac{(1-k)}{3}=2.
\end{eqnarray*}
In the above derivation, we used the identity
$$\sum_{i=m}^{\infty}\frac{i}{2^i}=\frac{(m+1)}{2^{m-1}}.$$
%\section{Densities Under the Paradoxical Map}
%Here we consider the evolution of densities under the map $\varphi(x)$ given in equation~(\ref{eqn:par}). Let us denote the Perron-Frobenius operator of this equation by $P$ and let $\rho_n$ be the $n$th iterate of a density under the dynamics of $\varphi(x)$. The Perron Frobenius operator is then given by~\cite{las-mac,las-mac2}
%\begin{equation}
%P\rho_n(x)=\frac{d}{dx}\int_{\varphi^{-1}([0,x])}\rho_n(u)\ud u.
%\label{app:pf1}
%\end{equation}
%For an initial density, $\rho_0(x)$, $\rho_n(x)=P^n\rho_0(x)$. We now note that
%\begin{equation}
%\varphi^{-1}([0,x])=[0,x/(1+x)]\cup[(2-x)/2,1].
%\label{app:inv}
%\end{equation}
%Substituting~(\ref{app:inv}) into~(\ref{app:pf1}) and performing the differentiation yields
%\begin{equation}
%P\rho_n(x)=\frac{1}{(1+x)^2}\rho_n\left(\frac{x}{1+x}\right)+\frac{1}{2}\rho_n\left(\frac{2-x}{2}\right).
%\label{app:pf2}
%\end{equation}
%At this point we note that 
%\begin{equation*}
%\rho_{n+1}(0)=\rho_{n}(0)+\frac{1}{2}\rho_{n}(1),
%\end{equation*}
%from which it follows that $\rho_{n+1}(0)\ge\rho_n(0)$. This implies that $\rho_n'(0)\ge0$. If we take $\rho_0(x)=1$, then we have $\rho_n'(0)>0$ for all $n$, in which case
%\begin{equation*}
%\lim_{n\rightarrow\infty}\rho_n(0)=\infty.
%\end{equation*}
\end{document}